\documentclass[11pt,a4paper]{article}

\usepackage[T1]{fontenc}
\usepackage[utf8]{inputenc}
\usepackage[spanish,english]{babel}
\usepackage[margin=2.5cm]{geometry}
\usepackage{amsmath,amssymb,bm}
\usepackage{booktabs}
\usepackage{array}
\usepackage{graphicx}
\usepackage{siunitx}
\usepackage{microtype}
\usepackage{xcolor}
\usepackage[numbers,sort&compress]{natbib}
\usepackage{hyperref}
\providecommand{\doi}[1]{\href{https://doi.org/#1}{\nolinkurl{doi:#1}}}
\usepackage{caption}
\usepackage{subcaption}
\usepackage{float}
\usepackage[section]{placeins}
\usepackage{needspace}

\graphicspath{{images/}}

\hypersetup{
  pdftitle={Data-driven assessment of spectral modes in restricted Boltzmann machines for magnetization reconstruction},
  pdfauthor={Diego Garcia Venegas, Joaquin Peralta, Claudia Loyola, Sergio Davis, Scott Broderick},
  colorlinks=true,
  linkcolor=blue!45!black,
  citecolor=blue!45!black,
  urlcolor=blue!45!black
}

\newcommand{\Tc}{T_{\mathrm c}}

\title{Data-driven assessment of spectral modes in restricted Boltzmann
machines for magnetization reconstruction}

\author{\begin{tabular}{c}
  Diego Garc\'ia Venegas$^{1}$, Joaqu\'in Peralta$^{1}$,
  Claudia Loyola$^{1}$,\\
  Sergio Davis$^{2}$, and Scott Broderick$^{3}$\\[0.5em]
  \small $^{1}$Universidad Andr\'es Bello, Santiago, Chile\\
  \small $^{2}$Comisi\'on Chilena de Energ\'ia Nuclear (CCHEN), Santiago, Chile\\
  \small $^{3}$University at Buffalo, Buffalo, United States
\end{tabular}}

\date{}

\begin{document}
\maketitle
\begin{abstract}
In this paper, we explore the impact of spectral mode selection on
magnetization reconstruction from Ising configurations. To address this
objective, we study 100 restricted Boltzmann machines (RBMs) on
$20\times20$ and $40\times40$ lattices at five temperatures
around the critical point, with 34\,000 spin configurations in each
dataset. A singular value decomposition of the trained RBM weights
identifies the leading spectral mode and the mode with the largest
contribution along the uniform magnetization direction. We weaken the
selected modes separately, with weight perturbations of equal Frobenius
norm within each network, and measure the error in the mean squared
magnetization reconstructed in one step from independent Ising configurations.
Paired block resampling estimates the uncertainty in the error difference
for each network. In the analysis, spectral amplitude and uniform
contribution select the same mode in 56\% of the RBMs. In all 44
remaining networks, weakening the largest uniform
contributor increases the absolute error more than weakening the leading
mode at both tested strengths. For data-driven modeling of magnetization,
the uniform contribution helps identify weight patterns that matter for
reconstruction even when their singular values are below the largest.
\end{abstract}

\section{Introduction}

In the design of magnetic materials, a basic question is how spin
interactions shape the temperature dependence of magnetization.
For example, simulations of Ising spin layers on a polarizable
substrate have been used to study this relationship in ultrathin iron
films on palladium \cite{strandburg1992}. The square-lattice Ising model
provides a simple setting for generating data on the relationship between
spin configurations and magnetization. Restricted Boltzmann machines
(RBMs) have been used to sample Ising configurations and recover
thermodynamic averages \cite{torlai2016,morningstar2018,dangelo2020}.
To understand the information captured by RBMs, we need to identify the
relationship between individual weight modes and the observable of
interest, here the mean squared magnetization. To accomplish this goal,
we systematically change individual singular modes of trained RBMs and
track the resulting reconstruction error.

The ferromagnetic Ising model on a square lattice has a known transition
temperature and spontaneous magnetization \cite{onsager1944,yang1952}.
Near the critical temperature, spin correlations extend across many
sites \cite{torlai2016}. The mean squared magnetization sums these
correlations over pairs of sites and includes contributions from
configurations of either magnetization sign. Principal component
analysis (PCA) of the spin covariance and second-moment matrices has
been used to identify collective patterns in these configurations
\cite{wang2016,hu2017,park2025}. These works provide a framework for
relating patterns in spin data to magnetic order. We build on this
approach by tracking the relationship between the singular modes of
the learned weights and the error in reconstructing magnetization.

For RBM weights, the largest singular value identifies the input direction
with the largest squared hidden-field response. Magnetization assigns
equal weight to all sites. A leading pattern with both positive and
negative regions may have a small projection on this uniform direction,
while a mode with a smaller singular value may make a larger contribution.
We therefore have two options for selecting a mode, either by spectral
amplitude or by uniform contribution. The reconstructed magnetization
also depends on the biases and activation probabilities of the RBM.
Spectral studies have identified structured patterns in RBM weights
and compared their spectra with those of the training data
\cite{decelle2017,decelle2018,funai2021}. Further studies have explored
filter sums, inverse-filter correlations and filter-informed one-step
reconstructions of Ising observables \cite{gu2022}. Information about
the sign of magnetization can also be distributed across hidden units
and concentrated by constraints during training
\cite{fernandez2023disentangling}.

In this paper, we study 100 RBMs of two different architectures, each
trained on Ising configurations at one of five temperatures. Contrastive
divergence with one step (CD-1) estimates parameter updates from differences
between data and one-step reconstruction statistics \cite{hinton2002,hinton2012}.
The reported protocol for generating the training data uses single-spin
Metropolis sampling with 60\,000 thermalization sweeps and 500 sweeps between retained
configurations. Each dataset contains 34\,000 configurations. We use
a singular value decomposition of the saved weights to identify the
largest contributor to the uniform response, then compare the effect of
weakening this mode with that of weakening the spectral leader. Within
each network we apply weight perturbations of equal Frobenius norm to
the selected modes and to three controls with little uniform contribution,
one mode at a time. We then reconstruct independent Ising configurations
in one step. {We assess the importance of
a mode by the change in the error of the reconstructed observable.}
This comparison tests which features of the learned weights matter for
magnetization reconstruction at the five temperatures studied.

\section{Model and methods}

\subsection{Ising configurations and RBMs}

We use the zero-field ferromagnetic Ising model on a periodic square
lattice of $N=L^2$ spins $s_i=\pm1$. Its Hamiltonian, magnetization per spin
and energy per spin are
\begin{equation}
 \mathcal H(\bm s)=-J\sum_{\langle i,j\rangle}s_i s_j,
 \qquad m=N^{-1}\sum_i s_i,\qquad e=\mathcal H/N.
 \label{eq:ising}
\end{equation}
Here $L$ is the number of sites on each side. We report energy per spin in
units of $J$ and temperature in units of $J/k_B$. The datasets use
$L\in\{20,40\}$ and
$T\in\{2.200,2.250,2.269,2.300,2.350\}$, around the infinite-lattice
critical value $k_B\Tc/J=2/\ln(1+\sqrt2)$ \cite{onsager1944}.
Finite lattices round the transition \cite{fisher1972finite}; the five
temperatures sample this region.

We flatten each spin configuration into a vector and encode its entries
as $v_i=(s_i+1)/2$. These $N$ binary visible variables are coupled to
$N_h$ binary hidden variables in a Bernoulli--Bernoulli RBM, with weights
$W\in\mathbb R^{N_h\times N}$ and biases
$\bm a\in\mathbb R^N$, $\bm b\in\mathbb R^{N_h}$:
\begin{equation}
 E(\bm v,\bm h)=-\bm a^{\mathsf T}\bm v-\bm b^{\mathsf T}\bm h
                -\bm h^{\mathsf T}W\bm v.
 \label{eq:rbm}
\end{equation}
The joint probability of visible and hidden states is proportional to
$e^{-E}$. This energy is dimensionless. The physical temperature specifies
the Ising data used to train a network. Connections run between the two
layers, so the states in one layer are independent when the other is fixed:
\begin{equation}
 \begin{gathered}
 p(h_j=1\mid\bm v)=g\!\left(b_j+\sum_iW_{ji}v_i\right),\qquad
 p(v_i=1\mid\bm h)=g\!\left(a_i+\sum_jW_{ji}h_j\right),\\
 g(z)=\frac{1}{1+e^{-z}}.
 \end{gathered}
 \label{eq:conditionals}
\end{equation}
{For an input $\bm v$, we first draw
the hidden binary state using the first conditional probability. The
second gives the probabilities of the reconstructed visible spins.
We use these probabilities to average visible observables analytically.}

The network is trained to represent the distribution of Ising configurations.
Contrastive divergence with one step (CD-1) estimates parameter updates
from differences between data statistics and one-step reconstruction
statistics \cite{hinton2002,hinton2012}. The reported training protocol
uses single-spin Metropolis sampling with 60\,000 thermalization sweeps and
500 sweeps between retained configurations, where a sweep comprises
$N$ attempted updates. Each dataset contains 34\,000 configurations,
with an 80/20 training/test split. The training settings are reported as
minibatches of 120, 2000 epochs and nominal learning rate $10^{-4}$.
The supplied optimizer resets its gradient moments at each minibatch.
The saved files contain weights and biases but do not identify the
optimizer revision used for each run. We evaluate the networks from their
saved parameters.

\subsection{Model ensemble and independent evaluation}
\label{sec:design}

There are ten saved RBMs for each size and temperature (Table~\ref{tab:design}).
The 100 networks were trained on separate datasets, with labels 000--009
within each size--temperature group and 3.4 million configurations in total.
These labels identify dataset/model pairs. The saved files do not record
the NumPy random states used for sampling and shuffling during training.
Both the lattice size and the hidden-to-visible ratio differ between
the two architectures.

\begin{table}[!htbp]
 \centering
 \caption{Model ensemble. $L$ is the lattice side, $N=L^2$ the number of
 visible spins, and $N_h$ the number of hidden units. Each size has ten saved
 models at each of five reduced temperatures, $k_BT/J=2.200,2.250,2.269,2.300,2.350$.
 All 100 models enter the spectral analysis and reconstruction comparison.}
 \label{tab:design}
 \begin{tabular}{rrrrrr}
 \toprule
 $L$ & $N$ & $N_h$ & $N_h/N$ & Models per temperature & Total\\
 \midrule
 20 & 400 & 256 & 0.640 & 10 & 50\\
 40 & 1600 & 1125 & 0.703125 & 10 & 50\\
 \bottomrule
 \end{tabular}
\end{table}

\begin{figure}[!htbp]
 \centering
 \includegraphics[width=\textwidth]{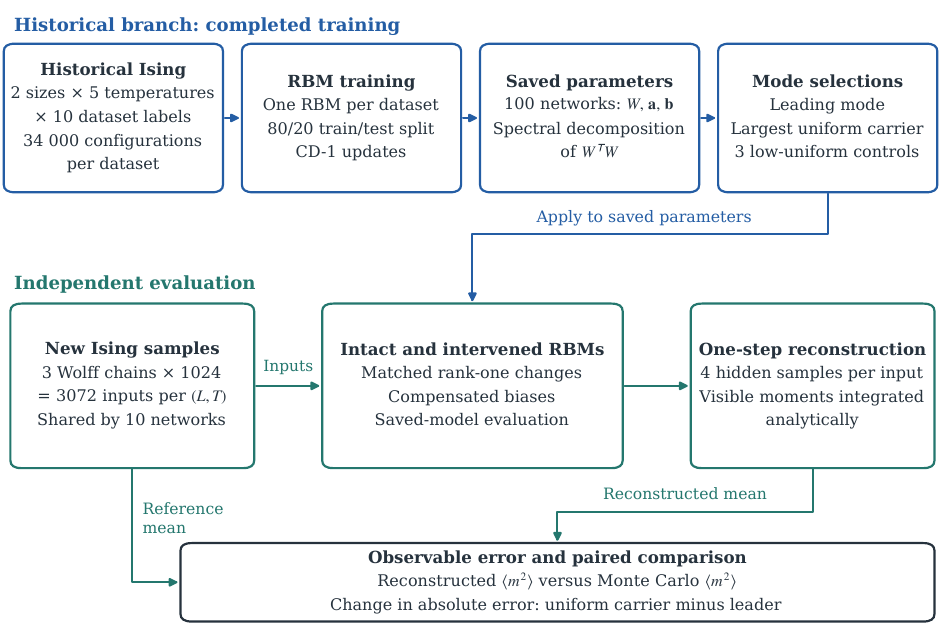}
 \caption{Spectral analysis and reconstruction test. The historical branch
 supplies 100 trained RBMs. Their weights determine the leader, largest
 uniform carrier and three low-uniform controls. Interventions weaken one
 mode at equal weight-change norm within a model and compensate the biases.
 The independent branch supplies 3072 Ising inputs per size--temperature
 cell. Intact and modified networks use the same inputs and hidden-sampling
 random numbers. One visible--hidden--visible step gives reconstructed means
 for comparison with the Monte Carlo reference. Evaluation starts from the
 saved parameters.}
 \label{fig:workflow}
\end{figure}

For evaluation, we generate independent Ising samples with Wolff
single-cluster updates \cite{wolff1989}. At each $(L,T)$, three chains
with explicit seeds start from ordered-up, ordered-down and random states.
After 5000 thermalization clusters, each chain supplies 1024 observations
separated by ten clusters. All ten RBMs in a size--temperature cell use
the same 3072 inputs, giving 30\,720 evaluation configurations across
the grid. The intact and modified versions of a model also use the same
random numbers for hidden sampling. Figure~\ref{fig:workflow}
summarizes the training archive and the evaluation procedure.

\subsection{Spectral modes and the magnetization direction}
\label{sec:geometry}

We decompose each saved weight matrix as
\begin{equation}
 W=\sum_{r=1}^{N_h}\sigma_r\bm u_r\bm v_r^{\mathsf T},\qquad
 C_W=W^{\mathsf T}W=\sum_{r=1}^{N_h}\lambda_r\bm v_r\bm v_r^{\mathsf T},
 \quad \lambda_r=\sigma_r^2.
 \label{eq:svd}
\end{equation}
Each mode pairs a unit visible vector $\bm v_r\in\mathbb R^N$ with a
unit hidden vector $\bm u_r\in\mathbb R^{N_h}$, so that
$W\bm v_r=\sigma_r\bm u_r$. We use all $N_h$ nonzero modes, ordered by
$\lambda_1\geq\lambda_2\geq\cdots$. The first is the leader.
Its visible vector maximizes $\|W\bm x\|^2$ among unit vectors $\bm x$.
For any input direction, this squared hidden-field response can be
calculated from the Gram matrix as
$\bm x^{\mathsf T}C_W\bm x=\|W\bm x\|^2$.

The total squared weight is
\begin{equation}
 \tau=\|W\|_F^2=\sum_{j,i}W_{ji}^2
      =\operatorname{tr}(W^{\mathsf T}W)=\sum_r\lambda_r,
 \label{eq:trace}
\end{equation}
where $F$ denotes the Frobenius norm. Since magnetization gives every
site the same weight, its direction is the unit vector
$\bm u_0=\bm1/\sqrt N$: projecting a spin configuration onto it gives
$\bm u_0^{\mathsf T}\bm s=\sqrt N\,m$.
{The vector $\bm u_0$ belongs to visible
space; the vectors $\bm u_r$ in Eq.~\eqref{eq:svd} belong to hidden space.}
We measure the alignment of the leader and its fraction of the total
squared weight by
\begin{equation}
 q_0=|\bm u_0^{\mathsf T}\bm v_1|^2,
 \qquad f_1=\lambda_1/\tau.
 \label{eq:leading}
\end{equation}
To account for every mode, expand
$W\bm u_0=\sum_r\sigma_r\bm u_r(\bm v_r^{\mathsf T}\bm u_0)$.
Orthogonality of the hidden vectors gives
\begin{equation}
 A_0=\frac{\|W\bm u_0\|^2}{\tau}
    =\frac{\bm u_0^{\mathsf T}C_W\bm u_0}{\tau}
    =\sum_r\underbrace{\frac{\lambda_r}{\tau}
             |\bm u_0^{\mathsf T}\bm v_r|^2}_{a_r}.
 \label{eq:a0}
\end{equation}
Each contribution $a_r$ combines squared amplitude with uniform alignment.
For $A_0>0$, its share of the uniform response and the cumulative share
through spectral rank $k$ are
\begin{equation}
 w_r=\frac{a_r}{A_0},\qquad
 \eta_1=w_1=\frac{f_1q_0}{A_0},\qquad F(k)=\sum_{r=1}^{k}w_r.
 \label{eq:shares}
\end{equation}
{The $w_r$ sum to one within the uniform
response. The trace fraction $f_1$ uses a different denominator, so a mode
can carry most of the uniform response and still account for little of
the total squared weight.}
We summarize concentration by
\begin{equation}
 n_{\mathrm{eff}}=\left(\sum_r w_r^2\right)^{-1}
 \label{eq:neff}
\end{equation}
which equals one when a single mode carries the uniform response and $n$
when $n$ modes share it equally. We calculate this quantity within each
network and then average over networks. A change of basis within a
degenerate subspace can change individual shares, while preserving their
sum and $A_0$.

Summing a row of $W$ gives the filter sum $S_j=\sum_iW_{ji}$, related to
the uniform response by
\begin{equation}
 \|W\bm u_0\|^2=\frac1N\sum_j S_j^2
   =\frac{N_h}{N}\left[\operatorname{Var}_j(S_j)+\overline S^{\,2}\right],
 \label{eq:filter-sums}
\end{equation}
where the variance is calculated with divisor $N_h$. Filter-sum
distributions, the variance of absolute filter sums and inverse-filter
correlations related to $W^{\mathsf T}W$ were studied by Gu and Zhang
\cite{gu2022}. Thus $A_0$ equals the second moment of the filter sums
multiplied by $N_h/N$ and divided by $\tau$.
Equation~\eqref{eq:a0} specifies how much each singular mode contributes.

\subsection{Physical observables and reference distributions}
\label{sec:references}

We compare reconstructions through $\langle m^2\rangle$. Squaring the
magnetization allows configurations of opposite sign to contribute without
cancelling. It also gives the sum of spin correlations over the lattice:
$\langle m^2\rangle=N^{-2}\sum_{i,j}\langle s_i s_j\rangle$.
For unit lattice spacing, the uncentered spin structure factor is
\begin{equation}
 S_s(\bm k)=\frac1N\left\langle\left|\sum_i
           s_i e^{-i\bm k\cdot\bm x_i}\right|^2\right\rangle,
 \qquad S_s(\bm0)=N\langle m^2\rangle.
 \label{eq:physical-target}
\end{equation}
The sum over the $N$ lattice wave vectors is
$\sum_{\bm k}S_s(\bm k)=N$. The fraction at zero wave vector is therefore
$S_s(\bm0)/N=\langle m^2\rangle$. We retain the spatial mean of each
configuration when calculating this quantity. The weight response along
the same uniform direction is $A_0$, normalized by the Gram trace $\tau$.
This puts the physical spin power and the weight response in a common
spatial basis.

Energy, magnetization and Fourier power are recalculated from all 3.4 million
historical configurations. For each Fourier shell, we divide its mean
power per wave vector by the mean power at zero wave vector. To estimate
the standard error, we recalculate the ratio after omitting each block
in turn from both means. Each block contains consecutive observations
from one trajectory, 1024 for the historical Metropolis data and 128 for
the Wolff data. We also use blocks of 64 to check the independent
evaluation. At every size and temperature, the historical
and independent means of $e$, $|m|$ and $m^2$ differ by at most 1.65
combined block standard errors.

To assess uniform alignment, we hold the eigenvalues of the visible
Gram matrix fixed and rotate its eigenbasis uniformly at random in
the $N$-dimensional visible space, including the $N-N_h$ null directions. Each squared overlap then has mean
$1/N$, giving
\begin{equation}
 \mathbb E A_0=\sum_r\frac{\lambda_r}{\tau}\frac1N=\frac1N.
 \label{eq:isotropic-mean}
\end{equation}
We report $NA_0$ to express the uniform response relative to this mean
and use 1999 random orientations per model.

\subsection{Mode selection, interventions and observable errors}
\label{sec:interventions}

The two selections are the leader ($r=1$) and the largest uniform carrier,
$r_*=\arg\max_r w_r=\arg\max_r\lambda_r|\bm u_0^{\mathsf T}\bm v_r|^2$.
They coincide in some networks. To select three controls with low uniform
contribution, we form a pool of modes with spectral ranks up to
$\min[N_h,\max(40,4r_*)]$, excluding the leader and carrier. We keep candidates with
uniform share at or below the median of this pool and select the three
closest to the carrier in $|\log(\sigma_r/\sigma_{r_*})|$.

For each selection, we attenuate one singular mode:
\begin{equation}
 W'_r=W-\kappa\bm u_r\bm v_r^{\mathsf T},\qquad
 \kappa=\delta\min_{r\in\mathcal R}\sigma_r,
 \quad \delta\in\{0.5,0.25\},
 \label{eq:intervention}
\end{equation}
Here $\mathcal R$ contains the leader, carrier and controls. We use
$\delta=0.5$ as the primary strength and repeat the comparison at $0.25$.
The outer product of two unit vectors has rank one and unit Frobenius norm,
so $\|W-W'_r\|_F=\kappa$. Within a network, the perturbations thus have equal
norm at each strength. The singular value of the selected mode determines
both the fraction removed, $\kappa/\sigma_r$, and the decrease in
trace, $2\kappa\sigma_r-\kappa^2$.

With $\Delta=W-W'_r$, we compensate the biases as
\begin{equation}
 \bm b'=\bm b+\tfrac12\Delta\bm1_N,\qquad
 \bm a'=\bm a+\tfrac12\Delta^{\mathsf T}\bm1_{N_h}.
 \label{eq:bias-compensation}
\end{equation}
With these biases, the hidden and visible fields keep their original
values at the binary centering point $v_i=h_j=1/2$. Equal norm refers
to the change in $W$. For the unnormalized uniform response
$R_0=\|W\bm u_0\|^2=\tau A_0$, reducing $\sigma_r$ to $\sigma_r-\kappa$
gives
\begin{equation}
 R_0-R'_0=R_0 w_r(2t_r-t_r^2),\qquad t_r=\kappa/\sigma_r.
 \label{eq:uniform-loss}
\end{equation}
For a carrier with $w_{r_*}>w_1$, the inequality
$\sigma_{r_*}\leq\sigma_1$ gives $t_{r_*}\geq t_1$ at fixed $\kappa$.
The factor $2t_r-t_r^2$ increases over the allowed interval $0\leq t_r\leq1$.

For an observable $O$, the reference is its Monte Carlo mean
$\langle O\rangle_{\rm MC}$. We compare this value with means reconstructed
by the intact and modified networks. For every physical input, we draw
four hidden states. We calculate the visible conditional moments for each
state and average them over hidden samples and inputs. Writing
$\mu_i=2p(v_i=1\mid\bm h)-1$, the conditional second moment is
\begin{equation}
 \mathbb E[m^2\mid\bm h]=
 \left(\frac1N\sum_i\mu_i\right)^2+
 \frac1{N^2}\sum_i(1-\mu_i^2).
 \label{eq:conditional-m2}
\end{equation}
The first term is the square of the conditional mean magnetization.
The second accounts for visible-spin fluctuations around their conditional
means. We evaluate the energy and Fourier power from the same conditional
spin products. Given $\bm h$, $\mathbb E[s_i s_j\mid\bm h]=\mu_i\mu_j$
for distinct sites, and $\mathbb E[s_i^2\mid\bm h]=1$.

We compare the absolute error of the observable mean and its change:
\begin{equation}
 \epsilon_O(W)=|\langle O\rangle_W-\langle O\rangle_{\rm MC}|,
 \qquad \Delta\epsilon_O^{(r)}=\epsilon_O(W'_r)-\epsilon_O(W).
 \label{eq:error-change}
\end{equation}
A positive $\Delta\epsilon_O^{(r)}$ means that the modification increases
the error relative to the intact network. We compare the carrier and
leader interventions through
\begin{equation}
 d_{m^2}=\Delta\epsilon_{m^2}^{(r_*)}-\Delta\epsilon_{m^2}^{(1)}
       =\epsilon_{m^2}(W'_{r_*})-\epsilon_{m^2}(W'_1).
 \label{eq:paired-contrast}
\end{equation}
{A positive $d_{m^2}$ means a larger error after
weakening the carrier than after weakening the leader. To determine whether
either intervention improves on the intact network, we examine the sign
of its own $\Delta\epsilon$.}
We also measure energy and the power fraction
$N^{-1}\sum_{\bm k\in\mathcal B}S_s(\bm k)$ in the nonzero
long-wavelength band $\mathcal B=\{\bm k:0<|\bm k|\leq\pi/5\}$.

We estimate evaluation uncertainty using 2000 paired block resamples.
Blocks of consecutive observations are formed within each chain and then
pooled. A resample uses the same blocks for the reference, the intact network
and every intervention. The resulting intervals describe evaluation error
for a fixed saved model. Variation between models is reported separately
as the sample standard deviation across the ten networks at each $(L,T)$.
The common physical inputs can correlate their evaluation errors.
The task remains one visible--hidden--visible reconstruction step from
independent physical data \cite{hinton2012}.

\section{Results}

\subsection{Uniform response outside the leading mode}

\begin{figure}[!htbp]
 \centering
 \includegraphics[width=\textwidth]{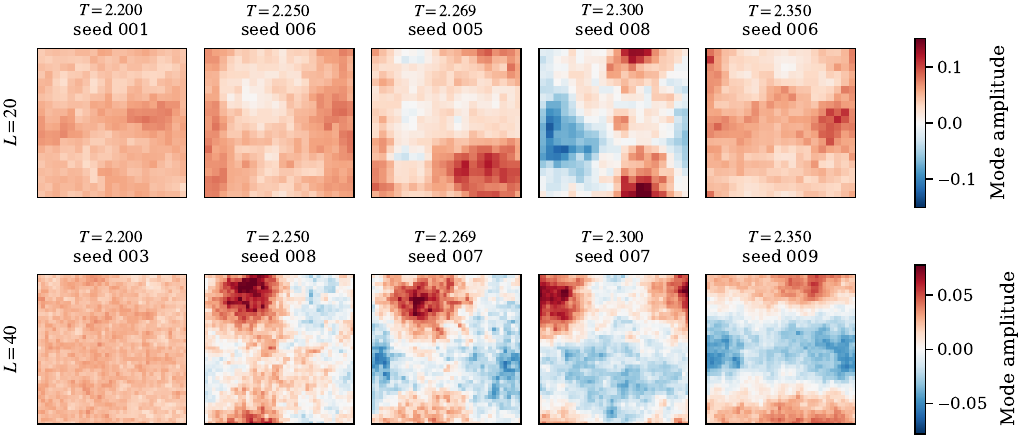}
 \caption{Leading visible eigenvectors of $C_W=W^{\mathsf T}W$ on the
 periodic lattices, $L=20$ above and $L=40$ below. Columns correspond
 to the five temperatures $k_BT/J$. For each $(L,T)$ we show the model
 whose $\lambda_1$ is closest to the median of the ten realizations.
 The label \texttt{seed} identifies its archived dataset. Each eigenvector
 has unit Euclidean norm and its sign gives a positive spatial mean.
 Red and blue indicate positive and negative amplitudes of the real-valued
 weight pattern. Each row has a common, symmetric color scale clipped at
 the largest of the panel-wise 99.5th percentiles of absolute amplitude.}
 \label{fig:modes}
\end{figure}

The leading patterns at $T=2.200$ are nearly uniform on both lattices
(Fig.~\ref{fig:modes}). At $L=40$, the modes displayed near $\Tc$ have
regions of both signs. For $L=20$, the largest departure from uniformity
is at $T=2.300$; the pattern at $T=2.350$ is again mostly uniform.
The examples in the figure are chosen by the leading eigenvalue closest
to the median within each size--temperature group.

\begin{figure}[!htbp]
 \centering
 \includegraphics[width=\textwidth]{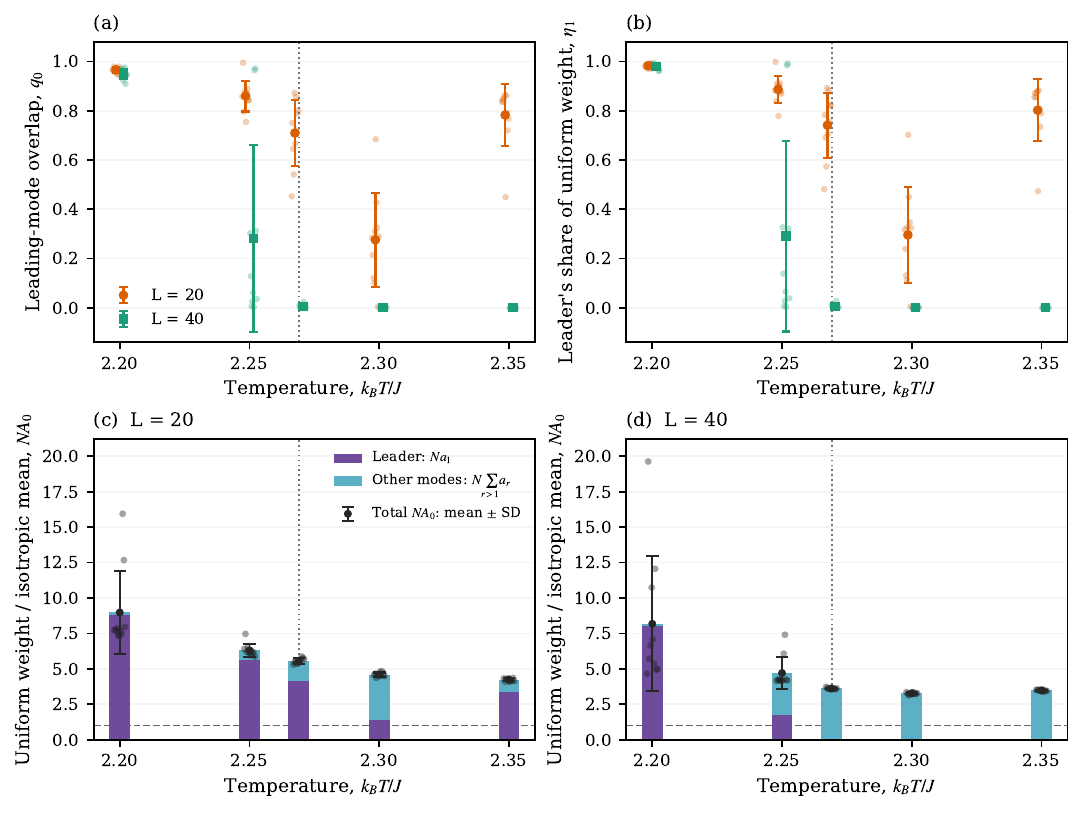}
 \caption{Leading alignment and partition of total uniform spectral weight.
 (a) Squared overlap $q_0$ with the uniform direction and (b) leading share
 $\eta_1$ for both sizes. Orange circles and green squares denote $L=20$ and $40$. Small points are
 individual models; large symbols and bars are means and sample standard
 deviations over ten models.
 (c,d) Total $N A_0$ split into the leading contribution $N A_0\eta_1$
 and the remaining contribution $N A_0(1-\eta_1)$ for each size.
 Purple and cyan stacks denote the leading and nonleading contributions,
 respectively, averaged after computing each contribution within each model;
 bars describe dispersion of the total. The horizontal reference is the
 isotropic mean $N A_0=1$. Vertical dotted lines mark the thermodynamic
 $\Tc$. Each temperature represents a separately trained ensemble.}
 \label{fig:partition}
\end{figure}

For the full $L=40$ ensemble, the mean leading overlap $q_0$ drops from
$0.9493\pm0.0211$ at $T=2.200$ to $0.00452\pm0.00745$ at $T=2.269$.
The uniform share $\eta_1$ drops from $0.98034\pm0.01035$ to
$0.00529\pm0.00884$ [Fig.~\ref{fig:partition}(a,b)]. Geometric error
bars here and below are sample standard deviations across ten models.

Over the same temperature interval, the total uniform response falls from
$NA_0=8.180\pm4.753$ to $3.6127\pm0.0466$. Even at $T=2.269$, each
model has a larger uniform response than any of its 1999 isotropic
surrogates, whose central 95\% ranges are approximately $0.93$--$1.08$.
Most of the measured response comes from modes below the leader.
Splitting $NA_0$ into its leading and remaining parts gives a decrease
from $8.0474$ to $0.0191$ in the first part and an increase from
$0.1326$ to $3.5936$ in the second [Fig.~\ref{fig:partition}(c,d)].

The relation differs at $L=20$. Between $T=2.300$ and $2.350$, the mean
$q_0$ rises from $0.276$ to $0.782$ while $NA_0$ falls from $4.598$ to
$4.223$. Here the leader becomes more uniform while the total response decreases.
Both the magnitude and the distribution of the uniform response vary
across temperature, with different trends in the two architectures.
Each temperature corresponds to separately trained networks.

\subsection{Distribution among subleading modes}

\begin{figure}[!htbp]
 \centering
 \includegraphics[width=\textwidth]{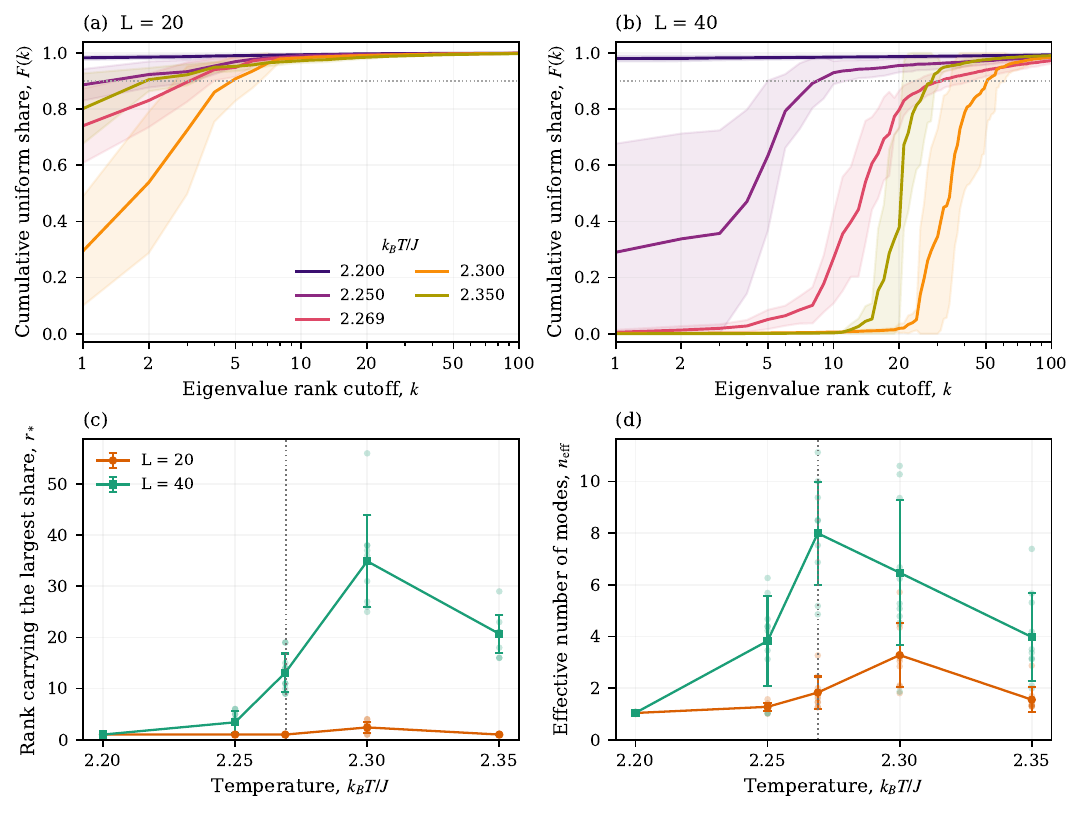}
 \caption{Distribution of uniform weight over spectral modes.
 In (a,b), $F(k)=\sum_{r\leq k}w_r$ gives the share carried by the first
 $k$ modes at $L=20$ and $40$. Color indicates temperature $k_BT/J$.
 Curves give the mean across ten models, with shading at one standard
 deviation and a dotted horizontal line at $F=0.9$. The logarithmic rank
 axis stops at 100. All 256 or 1125 nonzero modes enter the normalization.
 (c) Rank of the largest uniform carrier and
 (d) effective number $n_{\mathrm{eff}}=1/\sum_r w_r^2$, each calculated
 per model. Faint points are individual models; symbols and bars show their
 mean and standard deviation. Orange circles and green squares distinguish
 $L=20$ and $40$; the vertical reference is $\Tc$. The temperature of greatest modal spread differs between the architectures.}
 \label{fig:modal}
\end{figure}

At $L=40$, the effective number of uniform carriers rises from
$n_{\rm eff}=1.041\pm0.022$ at $T=2.200$ to $8.00\pm1.99$ at $T=2.269$
(Fig.~\ref{fig:modal}). At the latter temperature, the largest modal share
is $0.267\pm0.038$, with individual values from $0.195$ to $0.321$.
The response is shared by several modes within each saved network.

The first 20 modes contain about $80\%$ of the uniform weight and the
first 50 about $94\%$ at $L=40$, $T=2.269$. The mean rank needed to reach
$90\%$ is $33.3\pm9.6$. The large carriers are interspersed with modes
of small uniform share when ordered by eigenvalue. For randomly oriented
bases with the same spectra, the effective counts are of order hundreds.
The learned response is spread across several
modes but remains concentrated compared with this reference.

At $L=40$, $n_{\rm eff}$ decreases to $6.47\pm2.82$ at $T=2.300$ and
$3.98\pm1.69$ at $T=2.350$, while the leading share remains small.
At $L=20$, the largest sampled mean is $3.28\pm1.24$ at $T=2.300$,
followed by $1.55\pm0.48$ at $T=2.350$. The greatest modal spread occurs
at different temperatures in the two architectures.

\subsection{Effect of mode attenuation on reconstruction}
\label{sec:functional}
\label{sec:archive-functional}

\begin{figure}[!htbp]
 \centering
 \includegraphics[width=\textwidth]{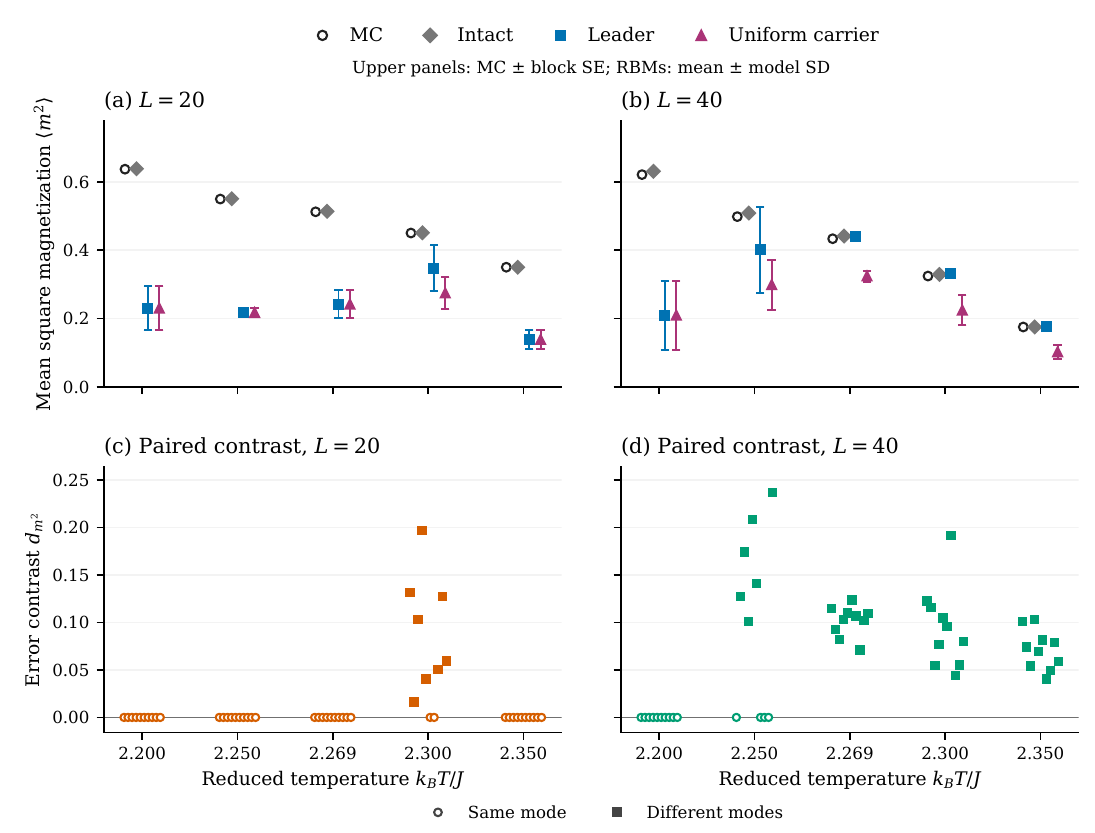}
 \caption{Reconstruction and paired mode comparison at $\delta=0.5$ for
 $L=20$ (left) and $40$ (right). Temperatures are in units of $J/k_B$.
 (a,b) $\langle m^2\rangle$ for the Ising reference (open black circles),
 intact RBMs (gray diamonds), leader attenuation (blue squares) and uniform
 carrier attenuation (purple triangles). The reference uses 3072 inputs
 per size--temperature cell, with block standard errors at block size 128.
 RBM symbols and bars show means and sample standard deviations across ten models.
 (c,d) Error contrast $d_{m^2}$ [Eq.~\eqref{eq:paired-contrast}] for all models:
 open circles are identical selections, filled squares distinct selections.
 Positive values mean a larger error under carrier attenuation. Conditional
 95\% intervals use 2000 paired block resamples within the three chains;
 many are smaller than the symbols. Horizontal offsets separate models.
 Weight changes have equal norm within a model, with compensated biases,
 four hidden samples per input and exact visible moments.}
 \label{fig:functional}
\end{figure}

The largest uniform carrier differs from the leader in 44 of the 100
networks. In every one of these 44, attenuating the carrier produces a
larger increase in the error of $\langle m^2\rangle$ than attenuating
the leader with the same weight-change norm (Fig.~\ref{fig:functional}).
At $\delta=0.5$, $d_{m^2}$ ranges from $0.0161$ to $0.2368$.
The other 56 networks select the same mode under both criteria and give
identical interventions. In all 100 networks, each low-uniform control
produces a smaller error increase than the carrier.

For $L=40$ at $T=2.269$, the carriers have spectral ranks 9--19.
The intact networks reconstruct $\langle m^2\rangle=0.4406$--$0.4428$,
close to the independent Ising mean $0.43401$. Attenuating the carrier
increases the absolute error by $0.0690$--$0.1255$. The leader changes
it by $-0.00716$--$0.00196$, and the absolute error change of every
low-uniform control is below $0.00154$. The carrier effect exceeds both
the initial reconstruction error and the effects of the comparison modes.

Coincident selections can also have a large effect on reconstruction.
At $L=20$, $T=2.200$, both criteria select the leader and their common
intervention increases the error by $0.406$ on average. At this size,
all eight distinct selections occur at $T=2.300$, with carrier ranks 2--4.
At $L=40$, the selections differ for six models at $T=2.250$ and for
all ten at each of $T=2.269,2.300,2.350$.

Reducing the strength to $\delta=0.25$ leaves all 44 contrasts positive,
with a minimum of $0.00814$. With blocks of 64 or 128 observations,
the lower bound of each paired interval is positive. The contrasts stay
positive when we use only the first halves of the chains and again when
we use only their second halves. The sign also persists if we keep the
reconstructed means fixed and use the historical Metropolis mean as the
reference in place of the Wolff mean.

In all 44 cases, weakening the carrier shifts the reconstructed mean
farther from the intact value than weakening the leader does. This holds
at both strengths, even though weakening the carrier removes less total
squared weight. The carrier/leader trace-loss ratio is $0.892$--$0.986$ at $\delta=0.5$ and
$0.905$--$0.988$ at $\delta=0.25$.

At $L=40$, $T=2.269$, weakening the carrier also increases the error in
energy. The increase ranges from $0.111$ to $0.181$ for the carrier and
from $0.0085$ to $0.0220$ for the leader. The nonzero long-wavelength power
can improve or worsen, depending on the model and intervention strength.

Selection by uniform contribution agrees with selection by alignment
alone in 99 of the 100 networks. In the exception ($L=40$, $T=2.250$,
label 003), alignment selects rank 8 and uniform contribution selects
rank 5. The reconstruction experiment compares rank 5 with the leader.
With only one disagreement between uniform contribution and alignment,
the ensemble mainly tests selection by physical direction against
selection by spectral amplitude.

\subsection{Weight response and spin fluctuations in Fourier space}
\label{sec:fourier-order}

The magnetization direction is the single Fourier mode $\bm k=0$.
The several contributions found above come from expressing this direction
in the learned eigenbasis. For a comparison with the physical spin
correlations, we use the Fourier basis and average the Gram matrix over periodic
translations:
$\overline C_W=N^{-1}\sum_{\bm a}R_{\bm a}^{\mathsf T}C_WR_{\bm a}$,
where $R_{\bm a}$ translates the lattice by $\bm a$.
The averaged matrix is diagonal in the normalized Fourier basis
$\phi_{\bm k}(\bm x)=N^{-1/2}e^{i\bm k\cdot\bm x}$,
with eigenvalues $D(\bm k)=\|W\phi_{\bm k}\|^2$ and
$\phi_{\bm0}=\bm u_0$. The average preserves $A_0$ and the trace and removes
off-diagonal terms between Fourier sectors.

\begin{figure}[!htbp]
 \centering
 \includegraphics[width=\textwidth]{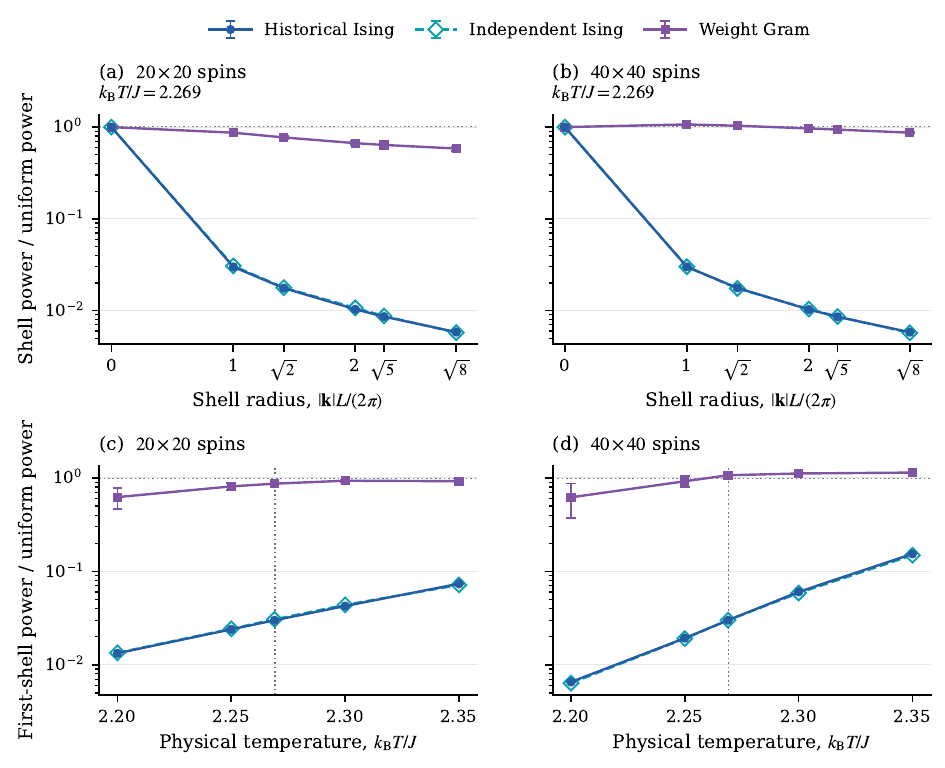}
 \caption{Fourier power of the Ising spins and Fourier response of the
 learned weights. Panels (a,b) compare $L=20$ and $40$ at $k_BT/J=2.269$.
 Points give the mean power per wave vector in a shell relative to the uniform sector.
 Shells group integer wave vectors by $n_x^2+n_y^2$, with
 $\bm k=2\pi(n_x,n_y)/L$.
 Panels (c,d) follow the first-shell ratio across temperatures for both
 sizes. At each size and temperature, blue circles represent 340\,000
 historical Metropolis configurations and open cyan diamonds 3072
 configurations from the independent Wolff runs. For the physical ratios
 $S_s(\bm k)/S_s(\bm0)$, bars show one standard error estimated by
 omitting one block at a time (1024 Metropolis or 128 Wolff observations
 from one chain). Purple squares show the mean of $D(\bm k)/D(\bm0)$
 across ten models. We calculate the ratio within each RBM, and the bars
 show the standard deviation across models. On the logarithmic vertical axes,
 dashed horizontal lines mark a ratio of one. Dotted vertical lines in
 (c,d) mark $T_c$. In this common Fourier basis the spin power peaks in the
 uniform sector and falls away much faster than the weight response.}
 \label{fig:structure-pilot}
\end{figure}

In all ten averaged matrices at $L=40$, $T=2.269$, the uniform sector
ranks ninth, below the first two nonzero shells. The maximum occurs
at $|\bm k|=2\pi/L$, with
$D(\bm0)/\max_{\bm k}D(\bm k)=0.9330\pm0.0119$.
The uniform sector ranks first at $T=2.200$ for every model and at every
temperature for $L=20$.

The physical structure factor has its maximum at $\bm k=0$ in all 100
historical trajectories (Fig.~\ref{fig:structure-pilot}). At $T=2.269$,
the first-shell power relative to $S_s(\bm0)$ is $0.02995\pm0.00009$
for $L=20$ and $0.02999\pm0.00011$ for $L=40$. Both uncertainties are
standard errors. At $L=40$
the learned weights give $1.069\pm0.013$ for the same ratio, where the
uncertainty is the standard deviation across models. The independent
Wolff data give the same ordering of the physical shells as the historical
data and show the same temperature dependence. In both physical datasets, power drops
much faster away from $\bm k=0$ than it does for the weight response.

The learned patterns also retain lattice organization. A common random
permutation of the visible sites preserves the spectrum and the uniform
response. Under this permutation the expected weight $D(\bm k)/\tau$
at each nonzero wave vector is $(1-A_0)/(N-1)$. The total weight in the
long-wavelength band $0<|\bm k|\leq\pi/5$ exceeds its permutation
expectation by factors of $3.12$--$4.94$.

\section{Discussion}

The intervention results relate the uniform projection of a weight mode
to its effect on magnetization reconstruction. In the 44 networks with
distinct selections, weakening the uniform carrier increases the error
of $\langle m^2\rangle$ more than weakening the leader at equal
weight-change norm. The carrier also produces a larger displacement of
the reconstructed mean itself, despite removing less total squared weight.
For this observable, the projection on the magnetization direction adds
information that the singular value alone does not provide.

In the larger architecture near $\Tc$, most of the uniform response comes
from subleading modes. When the two selections differ,
Eq.~\eqref{eq:uniform-loss} shows that the same attenuation removes more
unnormalized uniform weight from the largest carrier than from the leader.
That calculation uses only $W$.
The reconstructed observable also depends on the saved biases, sigmoid
probabilities and physical inputs, and we measure its error after passing
through the full network. In the Fourier comparison, $D(\bm k)$ measures
the squared linear field response and the spin correlations are averages
over the visible distribution.

Gu and Zhang connected filter statistics with thermodynamics and used
them in one-step reconstructions \cite{gu2022}. Magnetization information
in hidden units has also been studied through disentanglement
\cite{fernandez2023disentangling}, and spectral analyses describe the
patterns learned in the weights \cite{decelle2017,decelle2018,funai2021}.
Related interventions include weight pruning for quantum-state
reconstruction \cite{golubeva2022pruning} and hidden-unit ablations
associated with staggered-pattern alignment in a quantum-RBM preprint
\cite{mummaneni2026hidden}. Our comparison acts on visible singular modes
and fixes the norm of the weight change within each network. Uniform
contribution and alignment alone select the same mode in 99 of the 100
networks. The comparison with spectral amplitude is therefore the one
supported by this ensemble. The same approach could be tested for
staggered magnetization by assigning opposite weights to the two
sublattices when selecting a mode, then measuring the change in the
reconstructed observable.

The interventions remove $21$--$25\%$ or $42$--$50\%$ of the carrier
amplitude in the distinct-selection cases, so the measured response is
finite. They change energy as well as magnetization; changes in nonzero
Fourier power vary with the model and strength. The paired intervals
estimate evaluation uncertainty for a saved network, with physical
references shared across networks. Our observable is reconstructed in
one step from physical inputs. Equilibrium generation also involves the
relaxation and mixing of RBM chains
\cite{decelle2021equilibrium,bereux2025}.

Both lattice size and hidden-to-visible ratio change between the two
architectures. Their temperature trends describe these particular
networks. The incomplete record of the historical optimizer also limits
what can be said about how the representations developed during training.
The intervention comparison uses the saved parameters directly.
Individual eigenvectors introduce a further sensitivity when eigenvalues
are close: at $L=40$, $T=2.269$, the relative adjacent gaps around the
carriers range from $0.00117$ to $0.00681$. Small matrix perturbations
can then produce substantial eigenvector changes \cite{yu2015}.
Our selections use the eigenbases of the saved matrices. For a fixed
group of modes with orthogonal projector $P$, their summed uniform
contribution is $\|WP\bm u_0\|^2/\tau$, independent of the basis chosen
within that subspace.

\Needspace{12\baselineskip}
\section{Conclusions}

We compared two criteria for selecting weight modes in 100 RBMs trained
on Ising configurations: spectral amplitude and uniform contribution.
The mode carrying the largest uniform contribution differs from the
leader in 44 networks. In all 44, weakening the carrier produces a larger
increase in the reconstruction error of $\langle m^2\rangle$ at both
tested equal-norm strengths. In the remaining 56 networks, both criteria
select the same mode. In the larger architecture near the critical
temperature, the uniform response is shared among several subleading modes. For this
one-step reconstruction task, the direction associated with magnetization
identifies modes with a larger effect on the observable than the spectral
leader whenever the two selections differ.

\paragraph{Data availability.}
The data supporting the results of this study are available from the
authors upon request.

\FloatBarrier
\bibliographystyle{unsrtnat}

\end{document}